# Strong Electron-Phonon Coupling in Magic-Angle Twisted Bilayer Graphene

Cheng Chen[1,2]*, Kevin P. Nuckolls[3,4,a]*, Shuhan Ding[5], Wangqian Miao[6], Dillon Wong[3,4], Myungchul Oh[3,4,b], Ryan L. Lee[3,4], Shanmei He[2], Cheng Peng[2], Ding Pei[1], Yiwei Li[7], Chenyue Hao[8], Haoran Yan[5], Hanbo Xiao[1], Han Gao[1], Qiao Li[1], Shihao Zhang[1], Jianpeng Liu[1], Lin He[8], Kenji Watanabe[9], Takashi Taniguchi[10], Chris Jozwiak[11], Aaron Bostwick[11], Eli Rotenberg[11], Chu Li[12], Xu Han[12], Ding Pan[12], Zhongkai Liu[1], Xi Dai[12], Chaoxing Liu[13], B. Andrei Bernevig[4,14,15], Yao Wang[5,16] †, Ali Yazdani[3,4], Yulin Chen[1,2] †

*[1] Laboratory for Topological Physics and School of Physical Science and Technology, ShanghaiTech University, Shanghai, 201210 P. R. China*

*[2]Department of Physics, University of Oxford, Oxford, OX1 3PU, United Kingdom*

*[3]Joseph Henry Laboratories, Jadwin Hall, Princeton University, Princeton, NJ, USA*

*[4]Department of Physics, Princeton University, Princeton, NJ, USA*

*[5] Department of Chemistry, Emory University, Atlanta, GA 30322, United States*

*[6]Materials Department, University of California, Santa Barbara, California 93106-5050, USA*

*[7]Institute for Advanced Studies (IAS), Wuhan University, Wuhan 430072, China*

*[8]Center for Advanced Quantum Studies, Department of Physics, Beijing Normal University, Beijing 100875, People's Republic of China*

*[9]Research Center for Functional Materials, National Institute for Materials Science, Tsukuba, Japan*

*[10]International Center for Materials Nanoarchitectonics, National Institute for Materials Science, Tsukuba, Japan*

*[11]Advanced Light Source, Lawrence Berkeley National Laboratory, Berkeley, CA 94720, USA*

*[12]Department of Physics, Hong Kong University of Science and Technology, Clear Water Bay, Hong Kong, China*

*[13]Department of Physics, the Pennsylvania State University, University Park, Pennsylvania 16802, USA*

*[14]Donostia International Physics Center, P. Manuel de Lardizabal 4, 20018 Donostia-San Sebastian, Spain*

*[15]IKERBASQUE, Basque Foundation for Science, Bilbao, Spain*

*[16]Department of Physics and Astronomy, Clemson University, Clemson, South Carolina 29631, USA*

*[a]Present Address: Department of Physics, Massachusetts Institute of Technology, Cambridge, MA, USA*

*[b]Present Address: Department of Semiconductor Engineering, Pohang University of Science and Technology (POSTECH), Pohang, Republic of Korea*

*These authors contributed equally to this work

† Emails: yao.wang@emory.edu, yulin.chen@physics.ox.ac.uk

**The unusual properties of superconductivity in magic-angle twisted bilayer graphene (MATBG) have sparked enormous research interest[1-13]. However, despite the dedication of intensive experimental efforts and the proposal of several possible pairing mechanisms[14-24], the origin of its superconductivity remains elusive. Here, utilizing angle-resolved photoemission spectroscopy with micrometer spatial resolution, we have revealed flat band replicas in superconducting MATBG, where MATBG is unaligned with its hexagonal boron nitride (hBN) substrate[11]. These replicas exhibit uniform energy**

**spacing, approximately 150 ± 15 meV apart, indicative of strong electron-boson coupling. Strikingly, these replicas are absent in non-superconducting twisted bilayer graphene (TBG) systems, either when MATBG is aligned to hBN or when TBG deviates from the magic angle. Calculations suggest that the formation of these flat band replicas in superconducting MATBG are attributed to the strong coupling between flat band electrons and an optical phonon mode at the graphene K point, facilitated by inter-valley scattering. These findings, although do not necessarily put electron phonon coupling as the main driving force for the superconductivity in MATBG, unravel the unique electronic structure inherent in superconducting MATBG, thereby providing crucial information for understanding the unusual electronic landscape from which the superconductivity is derived.**

Magic-angle twisted bilayer graphene (MATBG) has attracted extensive research interest due to its remarkable tunability, offering a versatile platform for the study of strongly correlated electronic phenomena[1-13]. Many novel electronic states have been discovered in this low-density electron system, such as superconductivity, strongly correlated insulating states, pseudogap phases, topological phases and orbital magnetism[12,13]. The intricate interplay of strong electronic Coulomb interactions, which can override kinetic energy in flat band systems, likely underlies many of these phenomena[12,13]. Yet, the origin of the unusual superconducting state remains enigmatic, with proposed pairing mechanisms ranging from strong electronic correlations[14-17], electron-phonon interactions[18-22], spin fluctuations[23], and skyrmions[24].

In recent decades, angle-resolved photoemission spectroscopy (ARPES) has emerged as an important tool for investigating quantum materials, enabling the direct visualization of the electronic structure in momentum space[25,26]. However, the modest dimensions (usually 1-10 μm) of two-dimensional material devices and the prevalent twist angle inhomogeneity in MATBG devices[27] pose challenges for conventional ARPES techniques due to their limited spatial resolution (usually 50-500 μm). Fortunately, recent advancements in high-throughput

X-ray optics[28] have empowered us to perform high-quality ARPES measurements with sub-μm spatial precision (μ-ARPES), making it well-suited for unravelling the intricate electronic structure in MATBG devices[29,30].

In this study, we utilize μ-ARPES measurements to investigate and compare the electronic structures of superconducting MATBG devices (where MATBG is unaligned with its hBN substrate) and non-superconducting twisted bilayer graphene (TBG) systems, both hBN-aligned MATBG and TBGs deviating from the magic angle. Remarkably, in superconducting MATBG devices (unaligned with hBN), a distinct set of replicas of the flat band emerges at higher binding energies. These replicas exhibit robust spectral intensity features that emulate the energy and momentum characteristics of the original flat band. Notably, these replicas manifest across the entire momentum range of the moiré Brillouin zone (BZ), maintaining a uniform energy separation of 150 ± 15 meV. In contrast, when subjected to the same experimental conditions, the non-superconducting TBGs – either hBN-aligned MATBG devices or TBGs deviating from the magic angle – do not exhibit flat band replicas. These experimental findings naturally suggest a strong connection (see Supplementary Information (SI), Section XIII for more discussions) between the microscopic mechanisms responsible for the formation of flat band replicas and the occurrence of superconductivity in MATBG.

The observation of replicated bands at higher binding energies in the single-particle ARPES spectrum goes beyond the predictions of non-interacting band theories and often indicates a strongly correlated origin[31-33]. Through a comprehensive approach involving theoretical analysis and nonperturbative many-body simulations[19,34], we found that in superconducting MATBG (hBN-unaligned) devices, these observed replicas can be attributed to the strong coupling between flat band electrons to a transverse optical phonon mode at the graphene K point, facilitated by an inter-valley scattering process (discussed below, in the SI, Section VIII-X and in Ref. 34). Remarkably, such coupling diminishes in non-superconducting TBG devices, including hBN-aligned MATBG and TBGs deviating from the magic angle (discussed in SI, Section XI-XII). In summary, our discoveries provide valuable insights into the intricate interplay between flat band electrons and the bosonic degrees of freedom within MATBG, shedding light on the complex electronic configurations from which superconductivity emerges.

In this study, we first utilized superconducting (hBN-unaligned) and non-superconducting (hBN-aligned) MATBG devices previously characterized in a prior investigation[11], in which the twist angles and superconducting/spectroscopic properties of these devices were validated through scanning tunnelling microscopy/spectroscopy (STM/STS) analyses (for a concise summary, see SI, Section I). Furthermore, we conducted measurements on non-superconducting TBG devices with twist angles that slightly deviate from the magic angle in order to enhance the breadth and depth of our comparative analysis.

## Measurement schematic and sample geometry

A schematic representation of the μ-ARPES measurement is depicted in Fig. 1a, where the Au pad is connected to the MATBG sample to ensure proper grounding and prevent charge accumulation during measurements. The optical images displaying the device geometries for hBN-unaligned and hBN-aligned MATBG samples can be seen in Fig. 1b(i) and Fig. 1c(i), respectively. Notably, the μ-ARPES technique can directly resolve the MATBG band structure, as evidenced in the photoemission intensity maps for each sample, which distinguishes the MATBG regions from the hBN substrate (Fig. 1b(ii)) or single-layer graphene regions (Fig. 1c(ii)).

While two of the MATBG devices we studied share a nearly identical twist angle of ~1.08°, an optimal value associated with the maximum superconducting transition temperature ($T_c$) in transport measurements[1,35], superconductivity only occurs in the hBN-unaligned device but not in hBN-aligned devices (see Ref. 11 and further details available in SI, Section I). This disparity could be attributed to the influence of distinct moiré potentials stemming from different forms and degrees of coupling between MATBG and its hBN substrate, as illustrated in Fig. 1b(iii) and Fig. 1c(iii). In the unaligned sample (Fig. 1b(iii)), the hBN substrate introduces a high-frequency spatial background to the intrinsic MATBG moiré potential due to the large angle (~8°) mismatch between the hBN and MATBG, which does not significantly alter the electronic structure of MATBG. Conversely, in the aligned sample (Fig. 1c(iii)), the moiré periodicity (~13 nm) introduced by the hBN substrate closely matches that of intrinsic MATBG (13.8 nm) due to its small twist angle (~ 0.5°) relative to the bottom graphene layer. In this aligned sample, the graphene-hBN moiré potential and the resulting $C_2$-symmetry-breaking have a profound influence on the electronic structure of pristine MATBG, as will be discuss later.

## Flat band replicas observed in SC-MATBG

In Fig. 2, we present ARPES data from two superconducting MATBG (SC-MATBG, hBN-unaligned) Devices A & B. The measurements encompass the vicinity of the K points of the individual graphene layers spanning several moiré BZs (Fig. 2a). A three-dimensional (3D) band structure map of Device A is depicted in Fig. 2b, and six representative band dispersion spectra across the moiré BZs are shown in Fig. 2c. Near the Fermi level ($E_F$), we observe the flat band of MATBG, which extends throughout the entire moiré BZ and aligns well with single-particle calculations, as reported previously[29,30]. Remarkably, we observe previously unresolved replicas of this flat band at higher binding energies (Fig. 2c), which exhibit similar bandwidth, momentum range, and spectral distribution as the original flat band.

To provide enhanced clarity, we have captured these features in a series of ARPES dispersion spectra, cutting across the moiré BZ at different momentum locations (refer to labels 'cut 1' ~ 'cut 6' in Fig. 2a, b). These observed replicas and the original flat band display a uniform energy separation of 150 ± 15 meV, with intensities diminishing rapidly towards higher binding energies (Fig. 2c-d and detailed in SI Section II). Similar replica features are also observed in other superconducting MATBG devices (See Fig. 2e and SI Section III for Device B; SI Section V for STS signatures on Device F, as well as SI Section IV for data from literature[30]). These observed replica features cannot be explained by typical band hybridization.

First, they consistently manifest at uniform binding energies, unlike in typical band crossing/hybridization processes in which the characteristic energy varies. Second, their presence is not restricted to high-symmetry momenta within the moiré Brillouin zone. Finally, but most importantly, these replicas are absent in non-superconducting TBG devices, a point that will be elaborated upon in the following.

## Absence of replicas in non-SC samples

To compare our measurements on superconducting MATBG devices (Device A & B), we also have investigated non-superconducting (Non-SC) TBG devices, including both hBN-aligned MATBG (Device C) and TBGs with twist angle slightly deviating from the magic angle (Device D & E). As illustrated in Fig. 3a-c, in the hBN-aligned MATBG we find no indications of the replica features, despite the similarity of its twist angle and strain to the hBN-unaligned MATBG devices shown in Fig. 2. Instead, the band structure of this sample exhibits characteristic signatures of band hybridization between the Dirac bands of the two graphene layers, further modified by the commensurate MATBG/hBN moiré potential (Fig. 1b(iii)). It is noteworthy that this distinct electronic structure observed in hBN-aligned MATBG, as compared to hBN-unaligned MATBG (for a more comprehensive analysis, refer to SI, Section VI), underscores the significant impact of the moiré-commensurate hBN substrate[36], giving rise to unique emergent phenomena, such as the quantum anomalous Hall effect (QAHE)[5]. Moreover, typical signatures of band hybridization are also evident in other non-superconducting TBG devices (Device D & E and also Ref. 29) without any discernible signatures of replica band features (Fig. 3d-e).

## Strong electron-phonon coupling in MATBG

The observation of replica bands at higher binding energies in single-particle ARPES spectra typically signifies the presence of strong electron-boson coupling in the system[31-33]. In TBG, the concurrent emergence of these replica flat bands and superconductivity further indicates a shared underlying mechanism. This connection closely parallels the observation of shake-off replica bands in FeSe/$SrTiO_3$[31,37,38], which likely arise from the strong coupling of electrons in monolayer FeSe to an optical phonon mode in $SrTiO_3$. This strong electron-phonon coupling (EPC) was proposed to account for the much-enhanced superconducting transition temperature in this composite system compared to FeSe alone[31,37,38]. Despite the presence of various bosonic modes in MATBG, few are consistent with our observations. For instance, collective spin fluctuations, which couple with electrons to form magnetic polarons, can be ruled out, as they would yield unevenly spaced replica bands[39]. Plasmons, which couple with electrons to form plasmarons, are also unlikely candidates because the energy separation of the resulting replicas would be well below 150 meV[40,41] owing to the low carrier density in intrinsic MATBG/hBN[1]. In contrast, phonons, which couple with electrons to form polarons, are expected to possess mode energies around 150 meV, matching the energy spacing of the observed replicas, and thus stand as the most plausible origin[31-33,37,38].

Given the negligible electronic kinetic energy of the flat bands in MATBG (approximately 10~15 meV) compared to the large mode energy (~150 meV), we initially examine the intrinsic phonon spectra of graphene, without considering the renormalization caused by polaronic dressing. To induce replica features in both moiré BZs, the relevant phonon modes should reside around the Γ or K (K') point in the original graphene BZ, allowing them to couple with the corresponding intra- or inter-valley electron-hole processes, respectively. In Fig. 4a, we emphasize three phonon branches (distinguished by green, red and blue colors) with energies around 150 meV near the K point, indicating their potential to mediate inter-valley EPC. Notably, our frozen-phonon calculations reveal that the in-plane transverse optical (iTO, red line) mode exhibits the highest EPC strength dominant over the other two (iLO, iLA) relevant modes (as detailed in Fig. 4b and SI Section VII & VIII). This substantial difference in EPC strength is attributed to the symmetry of flat-band wavefunctions and two-center approximation[34], consistent with previous findings[42,43].

The iTO phonon mode scatters flat-band electrons between the moiré BZs at K and K' points (depicted in Fig. 4c), resulting in a series of evenly-spaced replica bands akin to those produced by forward-scattering phonons in each moiré BZ. The observed ARPES spectra can be well reproduced (Fig. 4d, with details in Method and SI, Section IX), with the energy separation determined by the phonon energy, which significantly surpasses the electronic bandwidth of ~15 meV [31-33].

Using the same EPC model, we further simulate the ARPES spectrum with the presence of valley-dependent potentials induced by the aligned hBN substrate. As shown in Fig. 4e, the polaronic replicas are significantly supressed with the presence of the hBN potential denoted as $\Delta$ (for details, see SI Section X), in agreement with our experimental observation. Additionally, to address the disappearance of polaronic replicas at larger and smaller twist angles in TBG, we conducted simulations to explore the relative replica intensity at varying bandwidths. As shown in Fig. 4f, the Poisson factor experiences a pronounced reduction as the bandwidth increases from the flat band limit ($w$=0) to finite values. Such a suppression reflects a similar mechanism to that influenced by the hBN potential $\Delta$, highlighting the competition between the single-particle electronic Hamiltonian and the valley-crossing EPC. This competition originates from the non-commuting nature of EPC matrices with the hBN potential or kinetic energy terms and reveals a universal phenomenon in materials with strong EPC (for detailed discussions, refer to SI Section X-XII.)

## Discussion and conclusions

We now discuss how this experimentally identified EPC may further contribute to electronic pairing in MATBG. Our raw estimation indicates that MATBG lies in the strong-coupling regime (for details and an explanation of strong coupling, see Section VIII of SI), consistent with the observation of a pseudogap above $T_c$[11]. We would like to emphasize that the involvement of phonons does not necessarily lead to conventional *s*-wave superconductivity, given the topological nature of multiple flat bands[34]. Previous studies have shown that inter-valley EPC can facilitate *d*-wave pairing in the intra-Chern-band channel[19,34,44], which opens the possibility of various pairing scenarios, including time-reversal breaking chiral *d*-wave pairing or gapless nematic *d*-wave pairing[45], with the latter consistent with prior STS observations of a nodal tunneling gap[11].

Due to the inter-valley nature of the EPC, a local Coulomb interaction in the chiral limit commutes with the EPC terms. As a result, the polaronic replicas induced by EPC are not affected by the presence of any local Coulomb interactions (see Section XII of SI). The coexistence of both strong EPC and Coulomb repulsions is anticipated to introduce further complexity to the pairing symmetry in MATBG and may collaborate in stabilizing superconductivity, as recently discussed in cuprates[46-48]. At low temperatures, these pairing properties would be reflected by electronic properties within the energy window of each replica. Unfortunately, this range lies beyond the reach of current μ-ARPES setup due to the temperature limitation (~ 20 K) and resolution constraints (~ 20 meV). However, the nature of electronic many-body states should not affect the quantification of polaronic replica features at the energy scale of 150 meV, thus is unlikely to substantially alter the interpretation presented above.

Our findings underscore the importance, although not necessarily the dominating role, of inter-valley phonon modes in superconducting MATBG, which act to dress the electronic states of the flat bands in ways that closely resemble those found in iron-based high-$T_c$ superconductors[31,37,38]. Going forward, we anticipate that forthcoming experimental endeavors will provide additional insights into the connection between the microscopic mechanisms driving replica flat bands and those underpinning superconductivity in twisted graphene systems. For instance, similar replica flat bands may emerge in superconducting magic-angle twisted trilayer graphene[49], while they may be absent in non-superconducting twisted monolayer-bilayer graphene[50]. Furthermore, we expect that the distinctive capabilities of μ-ARPES (and more advanced Nano-ARPES) will find widespread use in investigating strong correlation effects in moiré systems.

## Methods:

### Sample fabrication

MATBG Devices A, B and C are the Device B, F and C from Ref. 11 (the full description of the fabrication procedure can be found in this reference). Briefly, devices were fabricated using a 'tear-and-stack' method[51] in which a single graphene sheet is torn in half by van der Waals interaction with hBN. The two halves are rotated relative to each other and stacked to form MATBG. Graphene, graphite (in the aligned device) and hBN are picked up with polyvinyl alcohol. Then, to flip the heterostructure upside down, the heterostructure is pressed against an intermediate structure consisting of polymethyl methacrylate/transparent tape/Sylgard 184, and the polyvinyl alcohol is dissolved via water injection. The heterostructure is then transferred to a $SiO_2$/Si chip with pre-patterned Ti/Au electrodes. Residual polymer is dissolved in dichloromethane, water, acetone, and isopropyl alcohol. These chips have been annealed in ultra-high vacuum at 170 °C overnight and 400 °C for 2 h in previous STM/STS measurements[11]. The other devices share the similar fabrication and preparation method.

### Spatial- and Angle-resolved photoemission spectroscopy (μ-ARPES)

Synchrotron-based μ-ARPES measurements were performed at Beamline 7.0.2 (MAESTRO) of Advanced Light Source (ALS), USA and Beamline 07U of Shanghai Synchrotron Radiation

Facility (SSRF), China. The samples were annealed in ultra-high vacuum at 300 °C for 3h and measured under ultra-high vacuum below $3\times10^{-11}$ Torr. The photon energy of the incident beam is 95eV and the measurement is performed at room temperature. Data was collected using R4000 analyser upgraded with deflectors (ALS) and DA30 analyser (SSRF). The incoming photon beam is focused down to 2 um spot size by using a capillary mirror[28], and 1 um spot size using Fresnel zone plate (SSRF). The total energy and angle resolutions were 20 meV and 0.1°, respectively.

Laplacian (sum of second partial derivatives) plots are used to highlight the non-dispersive bands, as presented in Fig. 2 and Fig. 3. These plots are generated by summing the second partial derivatives of the original ARPES spectra along both axes in pixel units, then transforming them into energy and momentum space.

**Model calculation**

*1. Estimation of Electron-Phonon Coupling*

We adopt the following non-interacting-electron tight-binding Hamiltonian to calculate the electronic structure and EPC of MATBG,

$$H = -\sum_{\mathrm{I}i\alpha,\mathrm{J}j\beta} t(\boldsymbol{R}_{\mathrm{I}i\alpha} - \boldsymbol{R}_{\mathrm{J}j\beta}) c^{\dagger}_{\mathrm{I}i\alpha} c_{\mathrm{J}j\beta} \tag{1}$$

where $c^{\dagger}_{\mathrm{I}i\alpha}$, $c_{\mathrm{J}j\beta}$ are creation and annihilation operators for the $p_{\mathrm{z}}$ orbital of $i\alpha$ carbon atom in the I-th moiré superlattice ($\alpha$, $\beta$ are joint indices for sublattices and layers). The impact of atomic coordinates on the hopping parameter $t$ is approximated by Slater-Koster formula, whose dependence on atomic coordinates sets the stage for the EPC estimation. The Slater-Koster parameters are specified in the SI Section VII.

To evaluate the coupling in the mini-BZ, we further project the coupling matrix using Truncated Atomic Plane Wave (TAPW) method[43,52]. Thus, a moiré phonon near the $\Gamma_{\mathrm{M}}$ point can be approximated by iTO/iLA/iLO phonons at graphene $K/K'$ points. (see SI Section VIII for details). We set the moiré phonon energy $\omega_0 = 150$ meV, the mass of carbon atoms $m_c = 2.0\times10^{-26}$kg, and the characteristic phonon length $l_p = \sqrt{\hbar/(2m_c\omega_0)} = 34.0$ mÅ. The TAPW electrons and projected EPC constitute the model Hamiltonian for the TBG system

$$H = -\sum_{\bar{\boldsymbol{k}}\sigma} \boldsymbol{c}^{(\mathrm{TAPW})\dagger}_{\bar{\boldsymbol{k}}\sigma} h_{\bar{\boldsymbol{k}}} \boldsymbol{c}^{(\mathrm{TAPW})}_{\bar{\boldsymbol{k}}\sigma} - \frac{1}{\sqrt{N_m}} \sum_{\bar{\boldsymbol{k}}\sigma\bar{\boldsymbol{q}}\nu} \boldsymbol{c}^{(\mathrm{TAPW})\dagger}_{\bar{\boldsymbol{k}}+\bar{\boldsymbol{q}}\sigma} M_{\bar{\boldsymbol{q}}\nu} \boldsymbol{c}^{(\mathrm{TAPW})}_{\bar{\boldsymbol{k}}\sigma} \left(a_{\bar{\boldsymbol{q}}\nu} + a^{\dagger}_{-\bar{\boldsymbol{q}}\nu}\right) + \sum_{\bar{\boldsymbol{q}}\nu} \omega_0 a^{\dagger}_{\bar{\boldsymbol{q}}\nu} a_{\bar{\boldsymbol{q}}\nu} \tag{2}$$

where $\boldsymbol{c}^{(\mathrm{TAPW})}_{\bar{\boldsymbol{k}}\sigma}$ is the column vector of electron annihilation operators with moiré momentum $\bar{\boldsymbol{k}}$ and spin $\sigma$ in the TAPW basis and $h_{\bar{\boldsymbol{k}}}$ is the electronic hopping matrix for a specific moiré momentum $\bar{\boldsymbol{k}}$. The size of its moiré mini-BZ is denoted as $N_{\mathrm{m}}$ to distinguish from the system size $N$. For folded phonon branches, $a_{\bar{\boldsymbol{q}}\nu}$ is the annihilation operator with moiré momentum $\bar{\boldsymbol{q}}$ and $\nu$ labels the index of branches. The average distance between phonons and electrons is small compared to the moiré length scale. Therefore, the coupling matrix $M_{\bar{\boldsymbol{q}}\nu}$ can be estimated by the momentum-independent $M_0$, which can be evaluated by the frozen-phonon calculations.

As the experimentally relevant electrons lie in low-energy flat bands, we further project the TAPW orbitals onto the four flat-band orbitals, which can be expressed as the projection operator $\boldsymbol{c}_{\bar{\boldsymbol{k}}\sigma}^{(\mathrm{flat})} = P_{\bar{\boldsymbol{k}}}^{\dagger} \boldsymbol{c}_{\bar{\boldsymbol{k}}\sigma}^{(\mathrm{TAPW})}$ . In the projected Hamiltonian, the EPC matrix becomes a 4 × 4-dimensional $\widetilde{M_{\bar{\boldsymbol{k}}}}(\bar{\boldsymbol{q}}) = P_{\bar{\boldsymbol{k}}+\bar{\boldsymbol{q}}}^{\dagger} M_{\bar{\boldsymbol{q}}} P_{\bar{\boldsymbol{k}}}$. Based on its numerical distribution (see SI Section VIII), we find that $\widetilde{M_{\bar{\boldsymbol{k}}}}(\bar{\boldsymbol{q}})$ can be approximated by $Q_{\bar{\boldsymbol{k}}+\bar{\boldsymbol{q}}} g_{\bar{\boldsymbol{q}}} \eta Q_{\bar{\boldsymbol{k}}}^{\dagger}$, where $\eta$ is a constant diagonal matrix. Thus, the projected Hamiltonian can be written in an EPC-diagonal basis

$$H_{\mathrm{flat-band}} = \sum_{\bar{\boldsymbol{k}}\sigma} \boldsymbol{d}_{\bar{\boldsymbol{k}}\sigma}^{\dagger} Q_{\bar{\boldsymbol{k}}}^{\dagger} \varepsilon_{\bar{\boldsymbol{k}}} Q_{\bar{\boldsymbol{k}}} \boldsymbol{d}_{\bar{\boldsymbol{k}}\sigma} - \frac{1}{\sqrt{N_m}} \sum_{\bar{\boldsymbol{k}}\sigma\bar{\boldsymbol{q}}\nu} g_{\bar{\boldsymbol{q}}} \boldsymbol{d}_{\bar{\boldsymbol{k}}+\bar{\boldsymbol{q}}\sigma}^{\dagger} \eta \boldsymbol{d}_{\bar{\boldsymbol{k}}\sigma} \left(a_{\bar{\boldsymbol{q}}\nu} + a_{-\bar{\boldsymbol{q}}\nu}^{\dagger}\right) + \sum_{\bar{\boldsymbol{q}}\nu} \omega_0 a_{\bar{\boldsymbol{q}}\nu}^{\dagger} a_{\bar{\boldsymbol{q}}\nu} \tag{3}$$

where $\boldsymbol{d}_{\bar{\boldsymbol{k}}\sigma} = Q_{\bar{\boldsymbol{k}}}^{\dagger} \boldsymbol{c}_{\bar{\boldsymbol{k}}\sigma}^{(\mathrm{flat})}$ and the diagonal matrix $\varepsilon_{\bar{\boldsymbol{k}}}$ represents the energy of the flat bands.

*2. ARPES Spectral Simulation for TBG*

The strong EPC for the flat-band electrons leads to nonperturbative polaronic dressing effects. We consider the Lang-Firsov transformation for the coupled Hamiltonian

$$U_{\mathrm{LF}} = e^{-\frac{1}{\sqrt{N_m}\omega_0} \sum_{\bar{\boldsymbol{R}}\sigma\bar{\boldsymbol{q}}\nu} e^{-i\bar{\boldsymbol{R}}\cdot\bar{\boldsymbol{q}}} g_{\bar{\boldsymbol{q}}} \boldsymbol{d}_{\bar{\boldsymbol{R}}\sigma}^{\dagger} \eta \boldsymbol{d}_{\bar{\boldsymbol{R}}\sigma} \left(a_{\bar{\boldsymbol{q}}\nu} - a_{-\bar{\boldsymbol{q}}\nu}^{\dagger}\right)} \tag{4}$$

where $\boldsymbol{d}_{\bar{\boldsymbol{R}}\sigma}$ is the real-space annihilation operator of electron in the EPC-diagonal basis. Due to the separation of energy scales for electrons and phonons, we can employ the polaron ansatz for the ground-state wavefunction $|\Psi_G\rangle = U_{\mathrm{LF}}^{\dagger} |\psi_e\rangle \otimes |\psi_{\mathrm{ph}}\rangle$ . Moreover, since both the transformed coupling strength $g_{\bar{\boldsymbol{q}}}$ and temperature are much less than the phonon energy $\omega_0 =$ 150 meV, we further assume that the $|\psi_{\mathrm{ph}}\rangle$ can be approximated by a vacuum state $|0_{\mathrm{ph}}\rangle$. Thus, the electronic part $|\psi_e\rangle$ is determined by an effective Hamiltonian

$$\langle 0_{\mathrm{ph}} | U_{\mathrm{LF}} H U_{\mathrm{LF}}^{\dagger} | 0_{\mathrm{ph}} \rangle$$

$$= \sum_{\bar{\boldsymbol{R}}\bar{\boldsymbol{R}}'\sigma} \boldsymbol{d}_{\bar{\boldsymbol{R}}\sigma}^{\dagger} \langle 0_{\mathrm{ph}} | h_{\bar{\boldsymbol{R}}\bar{\boldsymbol{R}}'}^{*} | 0_{\mathrm{ph}} \rangle \boldsymbol{d}_{\bar{\boldsymbol{R}}'\sigma} - \sum_{\bar{\boldsymbol{R}}\bar{\boldsymbol{R}}'\sigma\sigma'\bar{\boldsymbol{q}}} \frac{N_v \left|g_{\bar{\boldsymbol{q}}}\right|^2}{N_m \omega_0} e^{i\left(\bar{\boldsymbol{R}}' - \bar{\boldsymbol{R}}\right)\cdot\bar{\boldsymbol{q}}} \left(\boldsymbol{d}_{\bar{\boldsymbol{R}}\sigma}^{\dagger} \eta \boldsymbol{d}_{\bar{\boldsymbol{R}}\sigma}\right) \left(\boldsymbol{d}_{\bar{\boldsymbol{R}}'\sigma'}^{\dagger} \eta \boldsymbol{d}_{\bar{\boldsymbol{R}}'\sigma'}\right) \tag{5}$$

The $h_{\bar{\boldsymbol{R}}\bar{\boldsymbol{R}}'}^{*}$ is the phonon-dressed electronic hopping matrix, whose specific form is shown in SI Section X. In the flat-band limit of TBG, this matrix is close to identity and, therefore, is irrelevant in determining the polaronic dressing. The ground state of the above equation determines the $|\psi_e\rangle$ in $|\Psi_G\rangle$.

The photoemission spectrum also involves excited states (denoted as $|\Phi\rangle$)

$$A(\bar{\boldsymbol{k}},\omega) = \mathrm{Im}\left\{\frac{1}{N_m}\sum_{\bar{\boldsymbol{R}}\bar{\boldsymbol{R}}'} e^{-i\left(\bar{\boldsymbol{R}}-\bar{\boldsymbol{R}}'\right)\cdot\bar{\boldsymbol{k}}} \sum_{\sigma\alpha,\Phi}\langle\Psi_G\left|c_{\bar{\boldsymbol{R}}\sigma\alpha}^{\dagger(\mathrm{flat})}\right|\Phi\rangle\langle\Phi\left|c_{\bar{\boldsymbol{R}}'\sigma\alpha}^{(\mathrm{flat})}\right|\Psi_G\rangle \frac{1}{\omega-i\Gamma+E_\Phi-E_G}\right\} \tag{6}$$

where $\Gamma$ is the Lorentzian broadening. With the aforementioned ground-state ansatz, the intensity of the *M*-th replica peak is explicitly determined as

$$\frac{1}{M!}\exp\left(-\sum_{\bar{\boldsymbol{q}}}\frac{N_v|g_{\bar{\boldsymbol{q}}}|^2}{N_m\omega_0^2}\right)\left(\sum_{\bar{\boldsymbol{q}}}\frac{N_v|g_{\bar{\boldsymbol{q}}}|^2}{N_m\omega_0^2}e^{i\bar{\boldsymbol{R}}\cdot\bar{\boldsymbol{q}}}\right)^M \tag{7}$$

It follows a Poisson distribution with the factor $\sum_{\bar{\boldsymbol{q}}}\frac{N_v|g_{\bar{\boldsymbol{q}}}|^2}{N_m\omega_0^2} \approx 0.11$, according to the frozen-phonon simulations. Focusing on the relative intensity of replica features and ignoring the interacting nature of electrons inside each replica, we produce the spectral simulation in Fig. 4d.

*3. Variational Non-Gaussian Ansatz for More General Models*

The Lang-Firsov transformation is suitable for the flat-band TBG model. The experiments presented in this paper also include TBG under more complicated conditions, including with hBN substrates and finite bandwidth away from the magic angle. These generalized cases can be modeled in the form of

$$H = \sum_{\bar{\boldsymbol{k}}\sigma}\boldsymbol{d}_{\bar{\boldsymbol{k}}\sigma}^\dagger h_{\bar{\boldsymbol{k}}}\boldsymbol{d}_{\bar{\boldsymbol{k}}\sigma} - \frac{1}{\sqrt{N_m}}\sum_{\bar{\boldsymbol{k}}\sigma}\boldsymbol{d}_{\bar{\boldsymbol{k}}+\bar{\boldsymbol{q}}\sigma}^\dagger M_{\bar{\boldsymbol{q}}}\boldsymbol{d}_{\bar{\boldsymbol{k}}\sigma}\left(a_{\bar{\boldsymbol{q}}}+a_{\bar{\boldsymbol{q}}}^\dagger\right) + \sum_{\boldsymbol{q}}\omega_{\boldsymbol{q}}a_{\bar{\boldsymbol{q}}}^\dagger a_{\bar{\boldsymbol{q}}} \tag{8}$$

where $h_{\bar{\boldsymbol{k}}}$ describes the band structure in general. The derivation and specific forms of Hamiltonian with hBN and bandwidth are detailed in the SI Sections X and XI.

It is important to note that the first term in Eq. (8) does not commute with the second term in this generalized case, making the Lang-Firsov transformation unsuitable for this case. To simulate the polaronic dressing in this generalized Hamiltonian, we employ a variational ansatz of the ground state $|\Psi_\mathrm{G}\rangle = U_\mathrm{NGS}^\dagger(\lambda)|\psi_\mathrm{e}\rangle|0_\mathrm{ph}\rangle$ with

$$U_\mathrm{NGS}^\dagger(\lambda) = \exp\left[\frac{\lambda}{\sqrt{N_m}}\sum_{\bar{\boldsymbol{R}}\sigma\bar{\boldsymbol{q}}v}e^{-i\bar{\boldsymbol{R}}\cdot\bar{\boldsymbol{q}}}\boldsymbol{d}_{\bar{\boldsymbol{R}}\sigma}^\dagger\eta\boldsymbol{d}_{\bar{\boldsymbol{R}}\sigma}\left(a_{\bar{\boldsymbol{q}}v}-a_{-\bar{\boldsymbol{q}}v}^\dagger\right)\right] \tag{9}$$

where $\lambda$ is the variational parameter in contrast to the fixed $g/\omega_0$ of the Lang-Firsov transformation for TBG. Due to the high phonon energy compared to any energy scales in Eq. (8), we still assume the post-transformation phonon state is vacuum. Therefore, the variational ansatz gives the total energy as a function of $\lambda$:

$$\begin{aligned} E_\mathrm{tot}(\lambda) &= \langle 0_\mathrm{ph}|\langle\psi_\mathrm{e}|U_\mathrm{NGS}(\lambda)HU_\mathrm{NGS}^\dagger(\lambda)|\psi_\mathrm{e}\rangle|0_\mathrm{ph}\rangle \\ &= E_\mathrm{kin} - \frac{N_v}{N_m}(2g\lambda-\omega_0\lambda^2)\sum_{\bar{\boldsymbol{R}}\bar{\boldsymbol{R}}'\sigma\sigma'\bar{\boldsymbol{q}}}e^{i\left(\bar{\boldsymbol{R}}'-\bar{\boldsymbol{R}}\right)\cdot\bar{\boldsymbol{q}}}\left\langle\psi_\mathrm{e}\left|\left(\boldsymbol{d}_{\bar{\boldsymbol{R}}\sigma}^\dagger\eta\boldsymbol{d}_{\bar{\boldsymbol{R}}\sigma}\right)\left(\boldsymbol{d}_{\bar{\boldsymbol{R}}'\sigma'}^\dagger\eta\boldsymbol{d}_{\bar{\boldsymbol{R}}'\sigma'}\right)\right|\psi_\mathrm{e}\right\rangle \end{aligned} \tag{10}$$

with the normalized kinetic energy $E_\mathrm{kin}$ follows the expression in the SI Sections X and XI. Unlike the analytically solvable Lang-Firsov transformation, the variational parameter $\lambda$ is obtained by numerical optimization of $E_\mathrm{tot}(\lambda)$. The self-consistent equations for different situations of the generalized Hamiltonian are derived in the SI Sections X and XI. This numerically determined $\lambda$ describes the relative strength of the polaronic dressing, where $\lambda$ reproduces $g/\omega_0$ when $H$ takes the flat-band TBG form in Eq. (3). The Poisson factor for the relative replica intensity is then obtained by $p = N_v\lambda^2$.

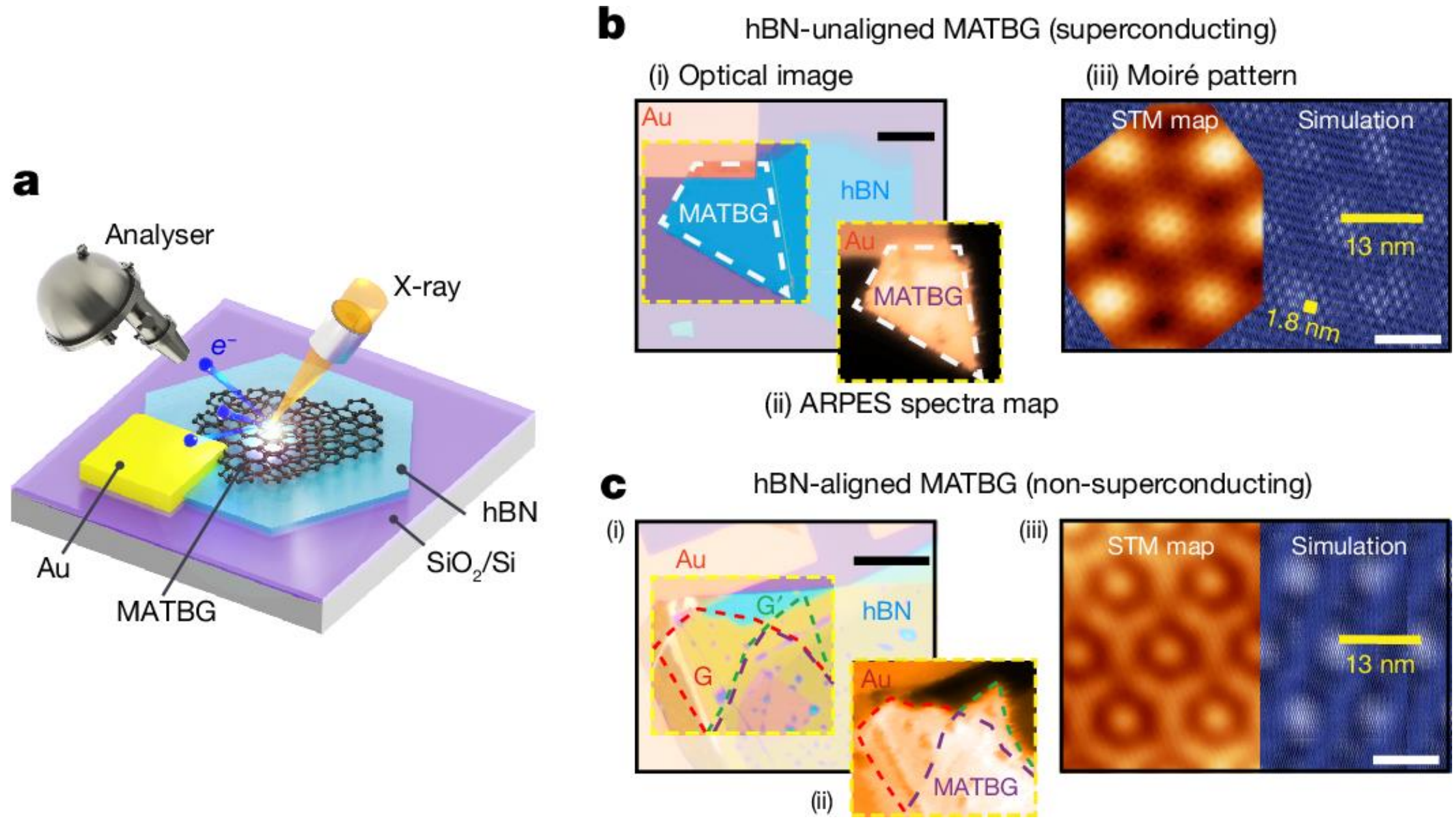

**Fig. 1| μ-ARPES measurement and MATBG device geometries. a**, Schematic diagram of the μ-ARPES measurement geometry. **b**, Basic characterization of the MATBG (twist angle 1.06°) Device A unaligned to its hBN substrate (8° rotational misalignment). (i) Optical image, with the substrate hBN, MATBG region (marked by white dashed lines) and the Au contact labelled. (ii) ARPES spectral intensity map covering the area enclosed by yellow dashed lines in (i) that highlights the MATBG region of the device. The spectral intensity is integrated within a binding energy of 0.2 eV, visualizing only conductive graphene and Au regions while omitting the hBN and $SiO_2$ regions. (iii) STM topographic image[10] ($V_b$ = −80 mV, $I$ = 300 pA) and a simulation of the hBN-unaligned MATBG lattice, showing two sets of moiré patterns with different periodicity: the larger one (~13.3 nm) from the graphene/graphene moiré and the smaller one (~1.76 nm, marked by the yellow bar) from the graphene/hBN moiré, respectively. **c**, Same as (b) for the MATBG (1.08°) Device C aligned with its hBN substrate (0.5° ± 0.1° rotational misalignment). (i) Optical image with monolayer graphene regions (G and G'; marked by red and green dashed lines) and the MATBG region (marked by purple dashed lines) labeled. The orange-colored regions of the hBN are the result of two overlapped hBN flakes used to encapsulate this sample's graphite back gate. (ii) ARPES spectral intensity map showing both monolayer graphene and MATBG regions corresponding to regions labelled in (i). (iii) STM topographic image[11] ($V_b$ = −300 mV, $I$ = 100 pA) and lattice simulation, showing a commensurate moiré structure from MATBG and graphene/hBN as both have periodicity around 13 nm.

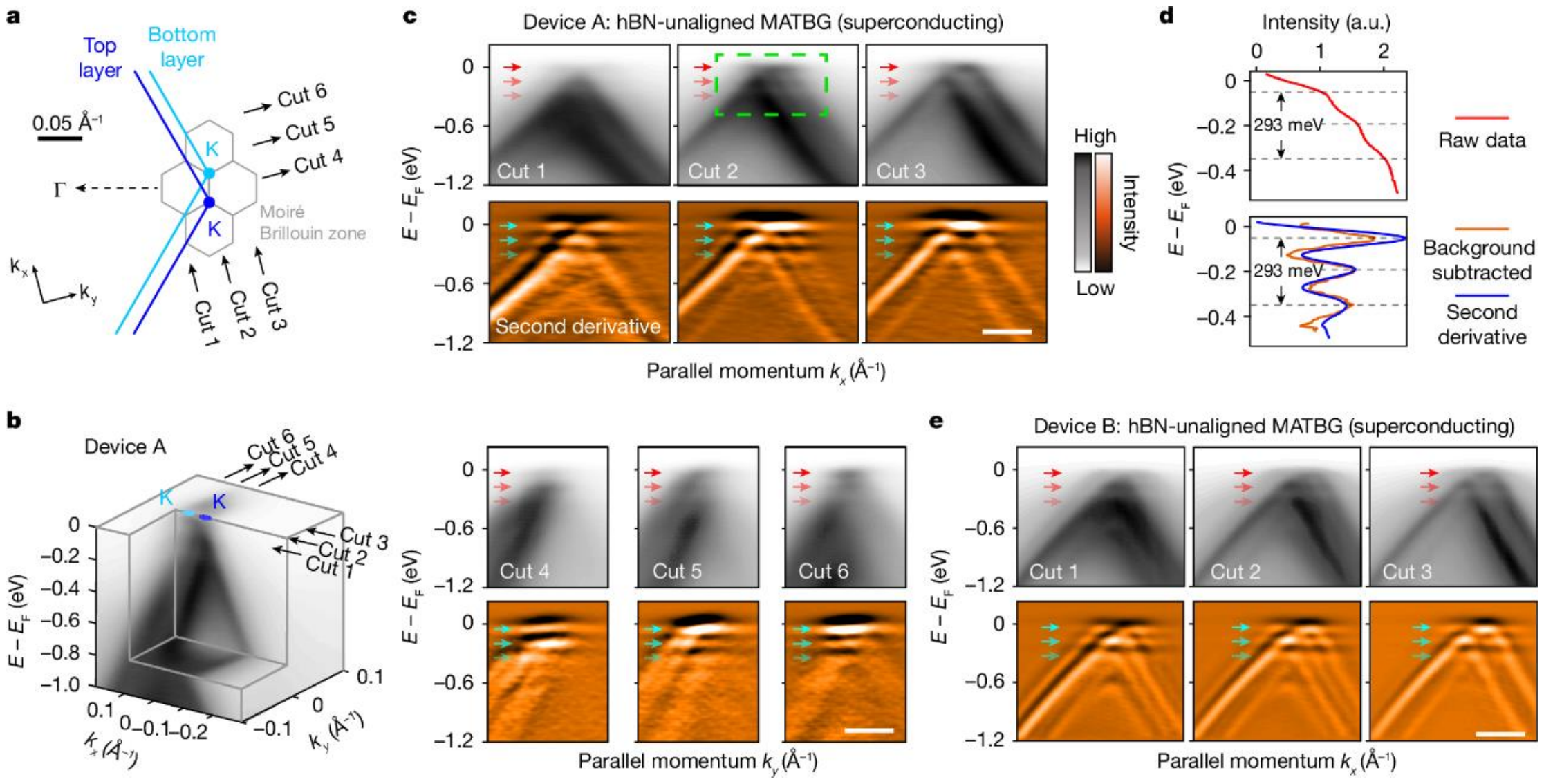

**Fig. 2| Observation of flat band replicas in superconducting MATBG (hBN-unaligned). a,** Illustration of the MATBG moiré Brillouin zones (BZs) around the K point of top and bottom monolayer graphene BZs. Grey lines show the moiré BZs of MATBG. 'cut1' ~ 'cut6' mark the coordinates of ARPES spectra in (c). **b,** 3D intensity plot of ARPES spectra in the vicinity of the graphene K point, presenting an overview of the band structure of MATBG. The ARPES measurements were conducted at room temperature (for details, see SI, Section III). **c**, ARPES dispersion plots taken on superconducting (SC) hBN-unaligned MATBG device (top rows, marked by 'cut1'~ 'cut6' in (a-b)) and their associated second partial derivative plots (bottom rows, see Method). Red / cyan arrows in top / bottom rows highlight the flat band and its first and second replicas. **d,** Top panel: Integrated ARPES spectra intensity in the green dashed line box in (c). Bottom panel: The red curve represents the same data shown in the top panel with the removal of a smooth background ($4^{th}$-degree polynomial). The blue curve represents the integrated intensity of the $2^{nd}$ derivative plot within the same green dashed line box. Both red and blue curves show an energy spacing between the flat band and replicas of 150 ± 15 meV. (Details can be found in Section II of SI). **e**, ARPES dispersion plots taken on a second superconducting hBN-unaligned MATBG, Device B, where flat band and replicas are observed (Details can be found in Section III of SI).

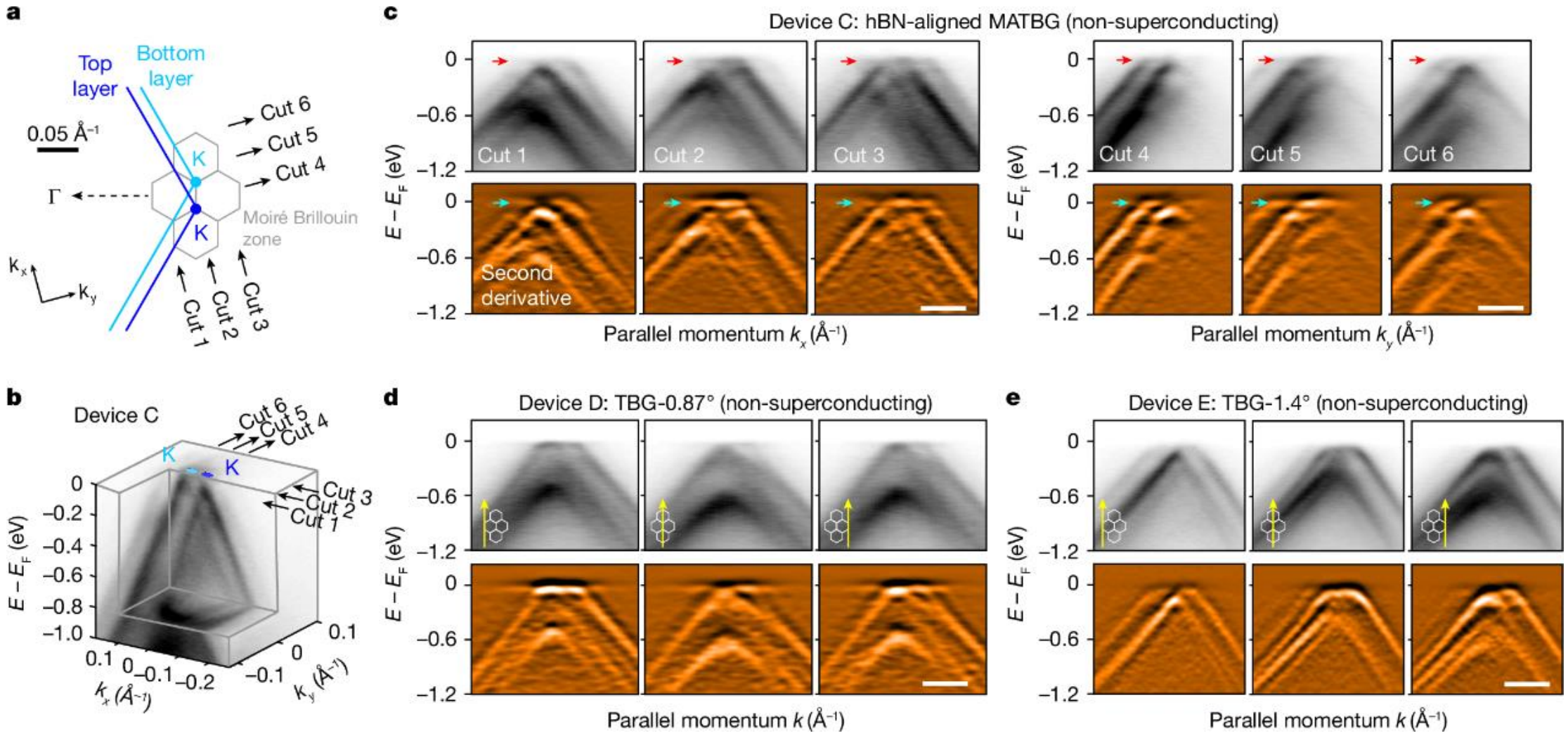

**Fig. 3| Absence of flat band replicas in non-superconducting TBG devices. a,** Illustration of the MATBG moiré BZs around the K point of top and bottom monolayer graphene. Grey lines show the moiré BZs of MATBG. 'cut1' ~ 'cut6' mark the coordinates of ARPES spectra in (c). **b,** 3D intensity plot of ARPES spectra in the vicinity of the graphene K point, presenting an overview of the band structure of hBN-aligned MATBG. Here, the electronic structure of MATBG is modified by the commensurate alignment of hBN as compared to the unaligned case illustrated in Fig. 2 (Detailed comparison can be found in Section VI of SI). **c**, ARPES dispersion plots and 2$^{nd}$ derivative plots taken along the same momentum directions (marked in (a-b)) as those in Fig. 2 for the hBN-unaligned MATBG sample. The flat band is labelled by red / cyan arrows. No signs of replicas of the flat band were observed, in contrast to the data shown in Fig. 2. **d-e**, ARPES dispersion plots (direction with respect to moiré BZs in (a) is illustrated) and 2$^{nd}$ derivative plots taken on the TBG devices with a twist angle of 0.87° and 1.4°, smaller/larger than the magic region. No signature of flat band replicas is evidenced.

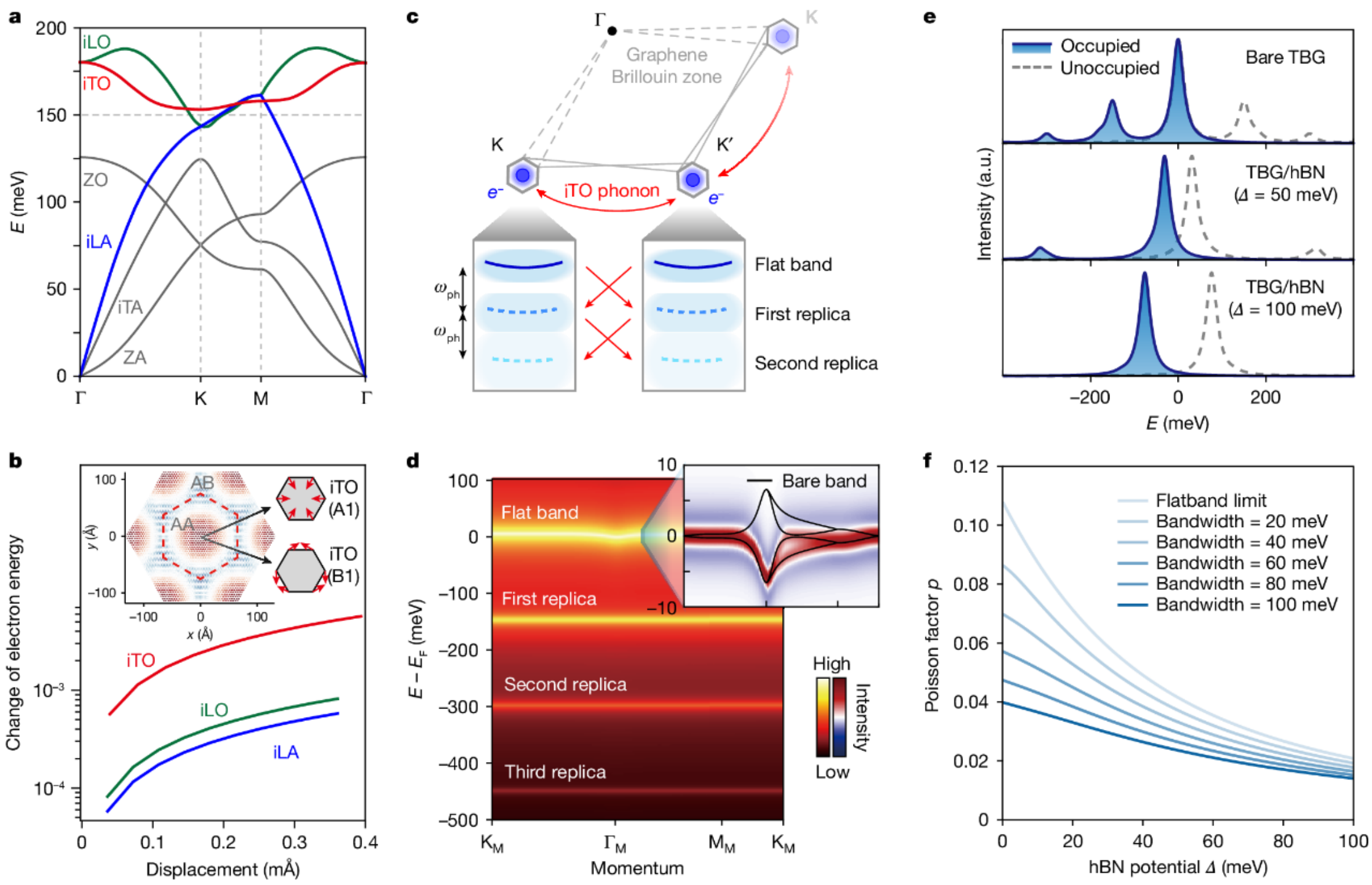

**Fig. 4| Inter-valley electron-phonon coupling in MATBG. a,** Calculated phonon spectra of single-layer graphene. Three phonon modes, in-plane longitudinal optical mode (iLO), in-plane transverse optical mode (iTO), in-plane longitudinal acoustic mode (iLA), with approximately 150 meV energy around the K point are highlighted. iTA is in-plane transverse acoustic mode, ZO is out-of-plane optical mode and ZA is out-of-plane acoustic mode. **b**, Calculated effective electron-phonon coupling (EPC) strength of iTO (in-plane TO), iLO and iTA phonon modes as a function of average lattice distortion. The EPC strength of iTO mode is almost an order of magnitude larger than the other two modes. Inset: Real-space lattice distortion magnitude induced by A1 and B1 modes of iTO phonons, which peaks at the AA-stacking regions and diminishes at the AB-stacking regions. (Details can be found in Section VIII of SI) **c,** Schematic diagram of the intervalley EPC in MATBG. Flat-band electrons from two moiré BZs at the K and K' points are exchanged via the scattering of iTO phonons, which generates shake-off replica bands separated by energy intervals of the iTO phonon energy. **d,** Simulated photoemission spectra of the flat band and replicas. The upper right panel shows the zoomed-in second-derivative spectra of the flat bands, where the solid lines are the calculated dispersion of bare bands without EPC (Details can be found in Section IX of SI). **e**, Spectral function obtained with the presence of band inversion induced by the hBN substrate potential Δ. The flat band replica is suppressed with increasing Δ. (SI, Section XII) **f**, The replica Poisson factor $p$, which reflects the relative intensity of ARPES replica, as a function of both hBN substrate potential Δ and the flat band bandwidth w. (SI, Section X and XI)